%% file: concat5.tex
\documentclass[aps,pra,twocolumn,groupedaddress]{revtex4-1}

\usepackage{amsfonts,amsmath,amsthm,amssymb,amsbsy}
\usepackage{graphicx,bbold}

\begin{document}


\title{Characterization of how errors accumulate in quantum computers}


\author{J. Gulliksen}

\author{D. D. Bhaktavatsala Rao}

\author{K. M\o lmer}
\affiliation{Department of Physics and Astronomy, Aarhus University, Ny Munkegade 120, DK-8000 Aarhus C, Denmark}
	

\date{\today}

\begin{abstract}
We study the achievements of quantum circuits comprised of several one- and two-qubit gates. Quantum process matrices are determined for the basic one- and two-qubit gate operations and concatenated to yield the process matrix of the combined quantum circuit. Examples are given of process matrices obtained by a Monte Carlo wavefunction analysis of Rydberg blockade gates in neutral atoms. Our analysis is ideally suited to compare different implementations of the same process. In particular, we show that the three-qubit Toffoli gate facilitated by the simultaneous interaction between all atoms may be accomplished with higher fidelity than a concatenation of one- and two-qubit gates.
\end{abstract}


\maketitle

\section{Introduction}
Since the first proposals were made to use quantum effects for computing purposes there has been a strong focus on how  errors and imperfections may harm and even prevent successful application of  quantum computing. A simple estimate suggests that if each single operation in a computation entails an error with a probability $p > 0$ then the application of $k$ operations will lead to a useful outcome with a probability that decreases exponentially $\sim (1-p)^k$. Error correction codes have provided a way to correct these errors up to a certain probability threshold, thereby allowing scalable, fault-tolerant quantum computing~\cite{PhysRevA.54.1098,royalcorrect}.

The error occurring in a single computational step such as a one- or two-qubit gate is often characterized by a single number, typically related to the overlap between the desired and actual output state, averaged over all input states. However, there is no guarantee that such a number encapsulates the accumulation of errors in a quantum circuit where the output state of one operation serves as the input to the next. Errors may build up coherently, so that error probabilities grow quadratically rather than linearly with time, or so that they compensate each other, cf., bang-bang control and composite pulses~\cite{PhysRevA.58.2733,Levitt198661,PhysRevA.87.052317}. Thus, a concatenation of two imperfect gates can lead to either unusable results or a correcting mechanism. To theoretically characterize a complete quantum circuit is a formidable task and is ultimately at odds with using a physical system to solve computationally hard problems. Still, a theoretical analysis of how errors propagate and accumulate in small systems may guide efforts to pick among different implementations of gates and assess optimal strategies for error correction.

In this article, we describe processes in a quantum system by the so-called $\chi$-matrices. In quantum computing we aim to implement definite gate  operations and process matrices account for the effects of error, e.g., due to dissipation and decoherence. It will be shown how $\chi$-matrices calculated once for one- and two-qubit gates can be concatenated to characterize circuits built from many of these gates. This will be exemplified in neutral atom quantum computing where the Rydberg blockade mechanism is used for two-qubit quantum gates~\cite{PhysRevLett.85.2208,RevModPhys.82.2313}. Circuits comprised of Rydberg mediated two-qubit gates may be directly compared to alternative multi-qubit implementations exploiting a single Rydberg atom's ability to simultaneously control a number of neighboring atoms.

The paper is organized as follows. In Sec.~\ref{sec:calcchi}, we review the definition of $\chi$-matrices and how they may be computed with Monte Carlo wave function simulations. In Sec.~\ref{sec:CG}, we describe how $\chi$-matrices for simple processes on few particles are concatenated to characterize large quantum circuits. In Sec.~\ref{sec:rydgte}, we introduce the Rydberg blockade gate scheme for quantum computing with neutral atoms. In Sec.~\ref{sec:TG}, we concatenate one- and two-qubit gate $\chi$-matrices in a neutral atom system to characterize the circuit performing a Toffoli gate. This we compare to a direct multi-atom Rydberg mediated implementation. In Sec.~\ref{sec:DC}, we conclude and present an outlook.

\section{\label{sec:calcchi}Process matrix identification}
Consider the action of a quantum process that takes an input density matrix $\rho$ describing a physical system with Hilbert space dimension $D$ to an output density matrix. Such a process is described as a completely-positive linear map $\mathcal{E}:\rho\to\mathcal{E}(\rho)$, and by introducing a complete basis of $D^2$ operators $\{E_n\}$ on the Hilbert space, $\mathcal{E}(\rho)$ can be written~\cite{jopchuang}
\begin{equation}
\mathcal{E}(\rho)=\sum_{mn}\chi_{mn}E_m\rho E_n^\dag\, .
\label{eq:opsum}
\end{equation}
The $D^2\times D^2$ elements $\chi_{mn}$ constitute the process matrix $\chi$.

Many techniques now exist to experimentally determine $\chi$. Standard quantum process tomography~\cite{nielsen2010quantum,jopchuang,PhysRevLett.78.390} successfully reproduces $\chi$ by measuring all output states via quantum \emph{state} tomography~\cite{PhysRevLett.105.150401,1464-4266-5-1-311}. This has been demonstrated in NMR~\cite{PhysRevA.64.012314,PhysRevA.67.042322}, optical~\cite{PhysRevLett.91.120402,PhysRevLett.93.080502}, and atomic systems~\cite{PhysRevA.72.013615}. Alternately, $\chi$ may be obtained making use of an ancillary system~\cite{PhysRevLett.86.4195,PhysRevLett.90.193601} or avoiding state tomography altogether through the use of suitable ``probe'' systems~\cite{PhysRevLett.97.170501,PhysRevA.75.062331,PhysRevA.75.044304}.

If the system is subject to known dissipation and decoherence mechanisms, the quantum system evolution may be modeled theoretically and the process matrix be calculated by solution of the quantum master equation. A gate operation typically involves application of time dependent laser pulses. Therefore, it is valuable to determine how losses and errors accumulate and contribute to different types of errors in the output. Such detailed studies may also serve to confirm the values of experimental parameters ~\cite{PhysRevA.85.042310,jcp.qptfourier}.

In a recent publication~\cite{PhysRevA.88.052129}, we described how to characterize a quantum controlled-phase gate subject to decay and dephasing. Instead of simulating the evolution of a complete set of $D^2$ input states we gain access to all elements of $\chi$ by evolving a single maximally entangled pure state of the system and an idle ancilla system of the same Hilbert space dimension~\cite{PhysRevLett.86.4195}. The system is propagated stochastically using the Monte Carlo wave function method, which on average reproduces results of a master equation evolution~\cite{PhysRevLett.68.580,Molmer:93,PhysRevA.45.4879}. Process characterization using this approach has a number of advantages: First, for large $D$, an adequate ensemble of wave functions is easier to store and evolve than density matrices. Second, obtaining $\chi$ through the output state data from an ensemble of wave functions is less costly, numerically, than from a density matrix~\cite{PhysRevA.88.052129}. Third, the stochastic evolution consists of a deterministic smooth evolution interrupted by "quantum jumps". Since useful quantum gates require excellent fidelity, jumps are rare and a single deterministic "no-jump" wave function suffices to provide a good estimate and rigorous bound on the process matrices describing the evolution~\cite{PhysRevA.88.052129}.

\section{\label{sec:CG} The process matrix for a quantum circuit}
Suppose the quantum circuit performing a computational task is composed of $N$ physical units. The Hilbert space of the entire system is then a tensor product of $N$ Hilbert spaces, each of dimension $d$. An implementation of a quantum process often requires using more than just the qubit states. However, since the physical units only process binary information we shall refer to them as qubits, even if we exploit states from a space larger than dimension $2$. On each qubit Hilbert space we assume the complete operator basis $\{e_{n_i}\}$. By merely forming tensor products of the basis operators, we obtain a complete operator basis $\{E_{n} = e_{n_1}\otimes \ldots\otimes e_{n_N}\}$ for the $N$ qubits, where the single index $n$ represents all values of the set $n_1,\ldots  n_N$.

If we assume that ({\it i}) process matrices $\chi$ correctly describe processes acting separately on one and two qubits of the circuit, and ({\it ii}) the decay and dissipation is independent and uncorrelated on different particles and at different times (no super-radiance or non-Markovian effects) then the application of several one- and two-qubit operations is exactly represented by an appropriate concatenation of the corresponding process matrices. The operator tensor product structure provides a convenient representation of the operators $E_m$ ($E^{\dag}_n$) in Eq.~(\ref{eq:opsum}) and enables a straighforward calculation of the $D^2\times D^2 =  d^{2N}\times d^{2N}$ dimensional process matrices $\chi$ for multi-qubit processes.

\subsection{Parallel concatenation}

\begin{figure}
\includegraphics[width=0.7\columnwidth]{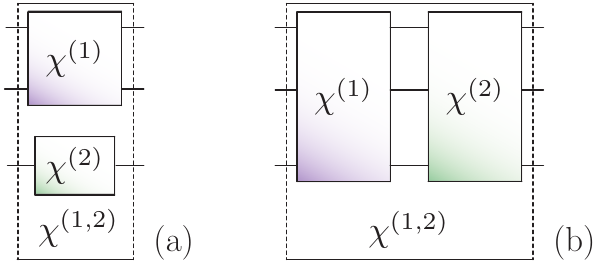}
\caption{\label{fig:catrules} Parallel and serial concatenation: Concatenation of (a) processes acting simultaneously on different qubits, and (b) processes acting sequentially on the same set of qubits. Expressions for the resulting process matrices $\chi^{(1,2)}$ are discussed in the text.}
\end{figure}
Suppose two subsystems are simultaneously subjected to processes independent of each other. These processes $\mathcal{E}^{(1)}$ and $\mathcal{E}^{(2)}$ may be described by the process matrices $\chi^{(1)}$ and $\chi^{(2)}$ respectively, illustrated as two- and one-qubit gates in Fig.~\ref{fig:catrules}(a). The combined three-qubit process matrix $\chi^{(1,2)}$ is simply the tensor product of the independent $\chi$ matrices. Other systems may be present but idle during the gate operation. They are acted on by the identity operator in the process matrix tensor product.

\subsection{Serial concatenation}
Most quantum algorithms make use of many computational steps, where the output of every step serves as the input to the subsequent one. In Fig.~\ref{fig:catrules}(b) we illustrate this situation for two consecutive three-qubit operations $\mathcal{E}^{(1)}$ and $\mathcal{E}^{(2)}$ characterized by $\chi^{(1)}$ and $\chi^{(2)}$ respectively. If the output $\mathcal{E}^{(1)}(\rho)$ of the input density matrix $\rho$ becomes the input of $\mathcal{E}^{(2)}$, what is the resulting $\chi$ matrix? Formally, the output of the sequential application of the operations is given by
\begin{equation}
\mathcal{E}^{(2)}\left(\mathcal{E}^{(1)}(\rho)\right)=\sum_{pq}\chi_{pq}^{(2)}E_p \left(\sum_{mn}\chi_{mn}^{(1)}E_m\rho E_n^\dag\right) E_q^\dag\, ,
\label{eq:concat1}
\end{equation}
Since the operators $E_r$ form a complete set, any product $E_pE_m$ can be expanded on these operators, that is, $E_pE_m=\sum_{mp}c_{pm}^r E_r$ and $E_n^\dag E_q^\dag=\sum_{nq}(c_{qn}^s)^\ast\, E_s^{\dag}$. Equation~(\ref{eq:concat1}) then becomes
\begin{equation}
\mathcal{E}^{(2)}\left(\mathcal{E}^{(1)}(\rho)\right)=\sum_{rs} \chi^{(1,2)}_{rs} E_r\rho E_s^\dag\, ,
\label{eq:concat2}
\end{equation}
where
\begin{equation}
\chi^{(1,2)}_{rs}=\sum_{mn,pq}c_{pm}^r\chi_{mn}^{(1)}\chi_{pq}^{(2)}(c_{qn}^s)^\ast\, .
\end{equation}
Note that although two consecutive processes may act on different subsets of some multi-qubit system both operations may be reformulated to act on the entire system through parallel concatenation.

It now becomes apparent that once the process matrices of all contributing gates in a circuit have been computed conclusively, we limit the cost of finding $\chi^{(1,2)}$ and thus of process matrices for larger quantum circuits. The assessment of how errors accumulate becomes a function of the width and depth of the quantum circuit.

\subsection{Example: Toffoli gate}
\begin{figure}
\includegraphics[width=\columnwidth]{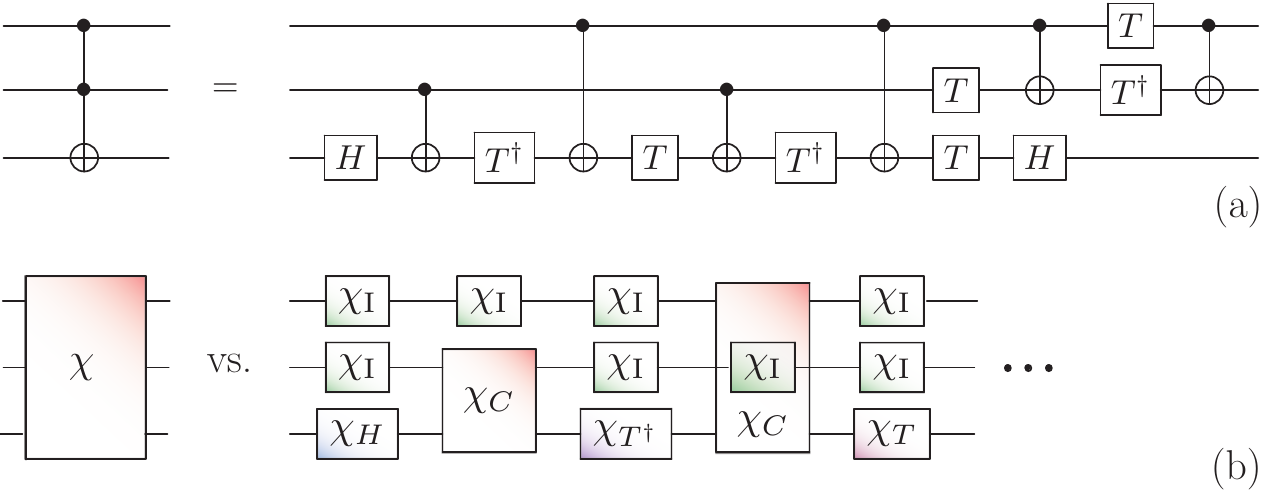}
\caption{\label{fig:toffcircuit} The Toffoli gate: (a) The three-qubit Toffoli gate on the left may be reproduced by a circuit of C-NOT, Hadamard ($H$), $T=\rm{exp}(i\pi\sigma_z/8)$ and $T^\dag$ gates, shown to the right. (b) The process matrix $\chi$ (left), characterizing the Toffoli gate may be calculated by concatenation of one- and two-qubit process matrices (right). The process matrices $\chi_C$, $\chi_H$, $\chi_T$ and $\chi_{T^{\dag}}$ characterize the C-NOT, Hadamard, $T$ and $T^\dag$ gates respectively. The ``identity'' process matrix $\chi_I$ indicates an idle qubit. }
\end{figure}

The Toffoli gate, or C$_2$-NOT gate, performs a controlled NOT operation on a target qubit based on the state of two control qubits. The Toffoli gate may be implemented as a sequence of six two-qubit C-NOT gates and nine one-qubit Hadamard and $T=\rm{exp}(i\pi\sigma_z/8)$ and $T^\dag$ phase gates, see Fig.\ref{fig:toffcircuit}(a). The gate and its generalization to higher numbers of control qubits (C$_k$-NOT) have applications as sub-modules in different quantum computing algorithms. Thus, it is relevant to determine the process matrix for its implementation in realistic systems.

In the analysis of the Toffoli gate process matrix we first simulated the propagation of quantum states through the sequence of one- and two-qubit gates in the full three-qubit Hilbert space. Such a calculation, e.g. using Monte Carlo wave functions to include dissipation, yields the full circuit process matrix $\chi_{\rm{cir}}$. Next, assuming the independence of errors occurring on different qubits and in different gates we apply the concatenation rules to obtain the circuit's process matrix $\chi_{\rm{cat}}$. Its repeated use of the same C-NOT $\chi$ matrix (cf. Fig.~\ref{fig:toffcircuit}(b)), which only needs a single calculation on a two-qubit system, attests to the advantage of the latter approach.

\section{\label{sec:rydgte} Rydberg blockade quantum gates}
A promising candidate for quantum computing involves neutral atoms held at closely spaced sites in far-off-resonance optical traps. The atoms may be individually addressed with laser fields and excited into high lying Rydberg states that feature strong, long distance dipole and van der Waals forces that can be used to mediate two-qubit interactions~\cite{PhysRevLett.85.2208,PhysRevA.72.022347,RevModPhys.82.2313}.

In Rubidium atoms, a convenient choice for the qubit states are the hyperfine ground states $|0\rangle\equiv|5s_{1/2},F=1,m_F=0\rangle$ and $|1\rangle\equiv|5s_{1/2},F=2,m_F=0\rangle$. They can be selectively excited to the Rydberg state $|r\rangle=|97d_{5/2},m_j=5/2\rangle$ by a two photon process using a 780-nm (480-nm) laser field, tuned by an amount $\Delta$ to the red (blue) of the intermediate $|p\rangle\equiv|5p_{3/2},F=3\rangle$ state. The Rabi frequency associated with the red (blue) detuned laser is $\Omega_R$ ($\Omega_B$), illustrated in Fig.~\ref{fig:cnot}(a). An atom that achieves excitation to the Rydberg state shifts the $|r\rangle$ state energy of all other atoms within the so-called blockade radius by an amount $\mathcal{B}$. Thus, one excited atom can prevent the resonant excitation of its neighboring atoms and this is the basis for effective quantum gates between them.

\begin{table}
\caption{\label{tab:params}Physical parameters for our simulations based on values discussed in Refs.~\cite{PhysRevLett.100.113003,saffman2011rydberg}.}
\begin{ruledtabular}
\begin{tabular}{lcr}
Experimental parameter & Symbol & Value \\ \hline
Detuning & $\Delta/2\pi$ & 2.0 GHz \\
Red Rabi frequency & $\Omega_R/2\pi$ & 118 MHz \\
Blue Rabi frequency & $\Omega_B/2\pi$ & 10 - 100 MHz \\
Rydberg blockade & $\mathcal{B}/2\pi$ & 20 MHz \\
Decay rate of $|p\rangle$ & $\gamma_p/2\pi$ & 6.07 MHz \\
Decay rate of $|r\rangle$ & $\gamma_r/2\pi$ & 0.53 kHz \\
Dephasing rate of $|r\rangle$ & $\gamma_d/2\pi$ & 1.0 kHz \\
\end{tabular}
\end{ruledtabular}
\end{table}

Dephasing of the Rydberg level normally associated with magnetic field noise and atomic motion is modeled by the operator $\hat{L}_{\gamma_d} = \sqrt{\gamma_d}(\mathbb{1}-2|r\rangle\langle r|)$, where $\gamma_d$ is the dephasing rate and $\mathbb{1}$ is shorthand for the identity operator. Spontaneous decay from a state $|y\rangle$ to a lower lying state $|z\rangle$ at a rate $\gamma_y$ is modeled by the jump operator $\hat{L}_{\gamma_y} = \sqrt{\gamma_y}|z\rangle\langle y|$. The effects of both dissipation mechanisms are simulated using the Monte Carlo wave function method~\cite{PhysRevA.88.052129}. Characteristic parameters are summarized in Table~\ref{tab:params}.

Adiabatic elimination by the effective operator formalism detailed in Ref.~\cite{PhysRevA.85.032111} provides a mechanism to decouple the intermediate optically excited state and describe the coherent and incoherent dynamics within the subspace of $|0\rangle, |1\rangle$ and $|r\rangle$. The system is then described by a Hamiltonian coupling a selected qubit state to $|r\rangle$ by an effective Rabi frequency. The formalism also provides an effective form for the operators describing decohering processes~\cite{PhysRevA.88.052129}.

\subsection{Rydberg blockade C-NOT gate}
In atomic quantum computing proposals single qubit gates amount to fast, resonant transitions within single atoms and can be made with high precision. Thus, for the purpose of this study we assume that the $\chi$ matrices associated with one-qubit gates are identical to the desired ones.  The two-qubit C-NOT gate depends on finite interactions between excited state atoms, lengthening gate time and making it prone to dissipation and decoherence.

\begin{figure}
\includegraphics[width=\columnwidth]{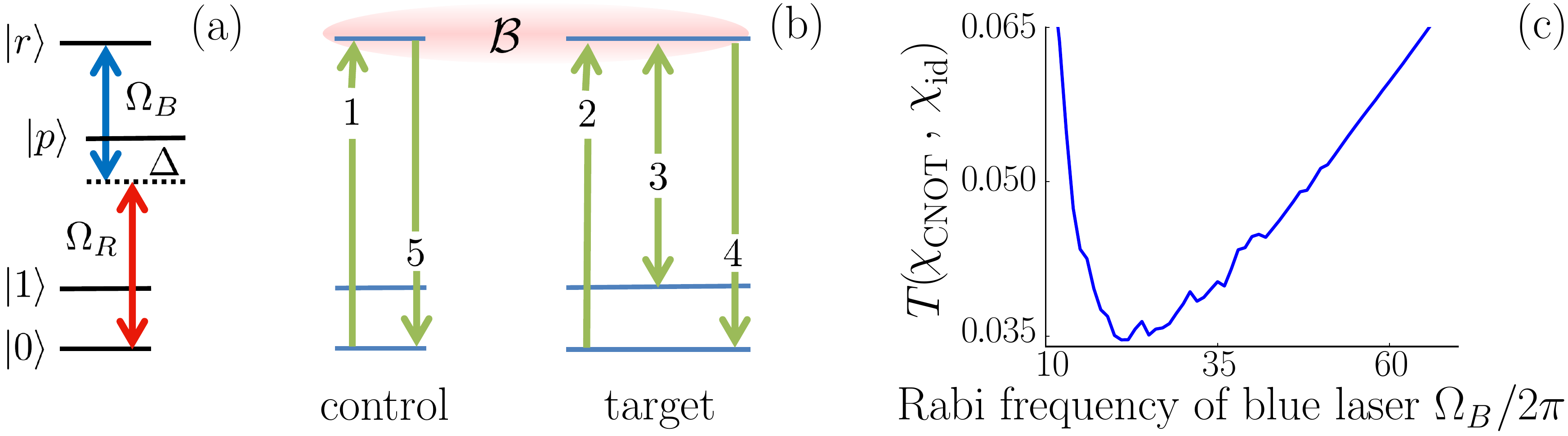}
\caption{\label{fig:cnot} Simulation and characterization of the C-NOT gate: (a) The red (lower) and blue (upper) laser fields drive $|0\rangle$, via $|p\rangle$, into the Rydberg state $|r\rangle$ by two-photon absorption. (b) Implementation of the C-NOT gate involves a sequence of two-photon $\pi$-pulses: In pulse 1 the control atom makes the transition $|0\rangle$ to $|r\rangle$. The target atom then makes the transitions $|0\rangle \leftrightarrow|r\rangle$ (pulse 2), $|1\rangle \leftrightarrow|r\rangle$ (pulse 3) and finally  $|0\rangle \leftrightarrow|r\rangle$ (pulse 4). In pulses 2-4 the target atom's states $|0\rangle$ and $|1\rangle$ are swapped, but only if the control atom is not in $|r\rangle$. Finally, in pulse 5 the control atom is driven from $|r\rangle$ back to $|0\rangle$. (c) Trace distance (see text in Sec.~\ref{sec:rydgte}A) between the ideal C-NOT process matrix and the process matrix calculated for the implementation shown in panel (b), with the parameters listed in Table~\ref{tab:params}. The trace distance is shown as a function of the blue laser Rabi frequency $\Omega_B$.}
\end{figure}

Figure~\ref{fig:cnot}(b) illustrates how a unitary C-NOT gate between two atoms can be implemented by a sequence of five perfect $\pi$-pulses. First transferring the control qubit's population from  $|0_c\rangle$ to $|r_c\rangle$ (pulse 1), then transferring the target qubit's population between $|0_t\rangle$ and $|1_t\rangle$ via the state $|r_t\rangle$ (pulses 2-4) and finally returning the control qubit's population from $|r_c\rangle$ to $|0_c\rangle$ (pulse 5). If the control qubit initially populates the state $|0_c\rangle$ the Rydberg blockade prevents any transfer during pulses 2-4. Thus, a NOT operation on the target qubit is conditioned on the control qubit initially populating the state $|1_c\rangle$, defining it to be a C-NOT operation.

Monte Carlo wave functions were used to simulate the five $\pi$-pulse implementation of the C-NOT gate (Fig.~\ref{fig:cnot}) with the parameters of Table~\ref{tab:params}. The performance of the gate was investigated as a function of the blue laser Rabi frequency $\Omega_B$. To provide a simple quantitative measure we applied the trace distance measure $T(\chi_{\rm sim},\chi_{\rm id})$ between the simulated and ideal process matrix, where  $T(A,B) \equiv\frac{1}{2}\lVert A-B\rVert_{tr}$ and $\lVert C \rVert_{tr}={\rm Tr}(\sqrt{C^{\dag}C})$ is the trace norm. Note that this distance measure is less ``forgiving'' than, for example, measures based on the trace overlap~\cite{PhysRevA.85.042310}. In Fig.~\ref{fig:cnot}(c) we show trace distance between a simulated C-NOT gate process matrix and the ideal, unitary process matrix. At low values of $\Omega_B$ the gate experiences greater dephasing errors from population in the Rydberg state, due to long gate times. At large $\Omega_B$ the blockade mechanism becomes inefficient. Thus, the optimum Rabi frequency lies between these two regimes.

\section{\label{sec:TG}The Toffoli gate by Rydberg blockade}
\begin{figure}
\includegraphics[width=0.98\columnwidth]{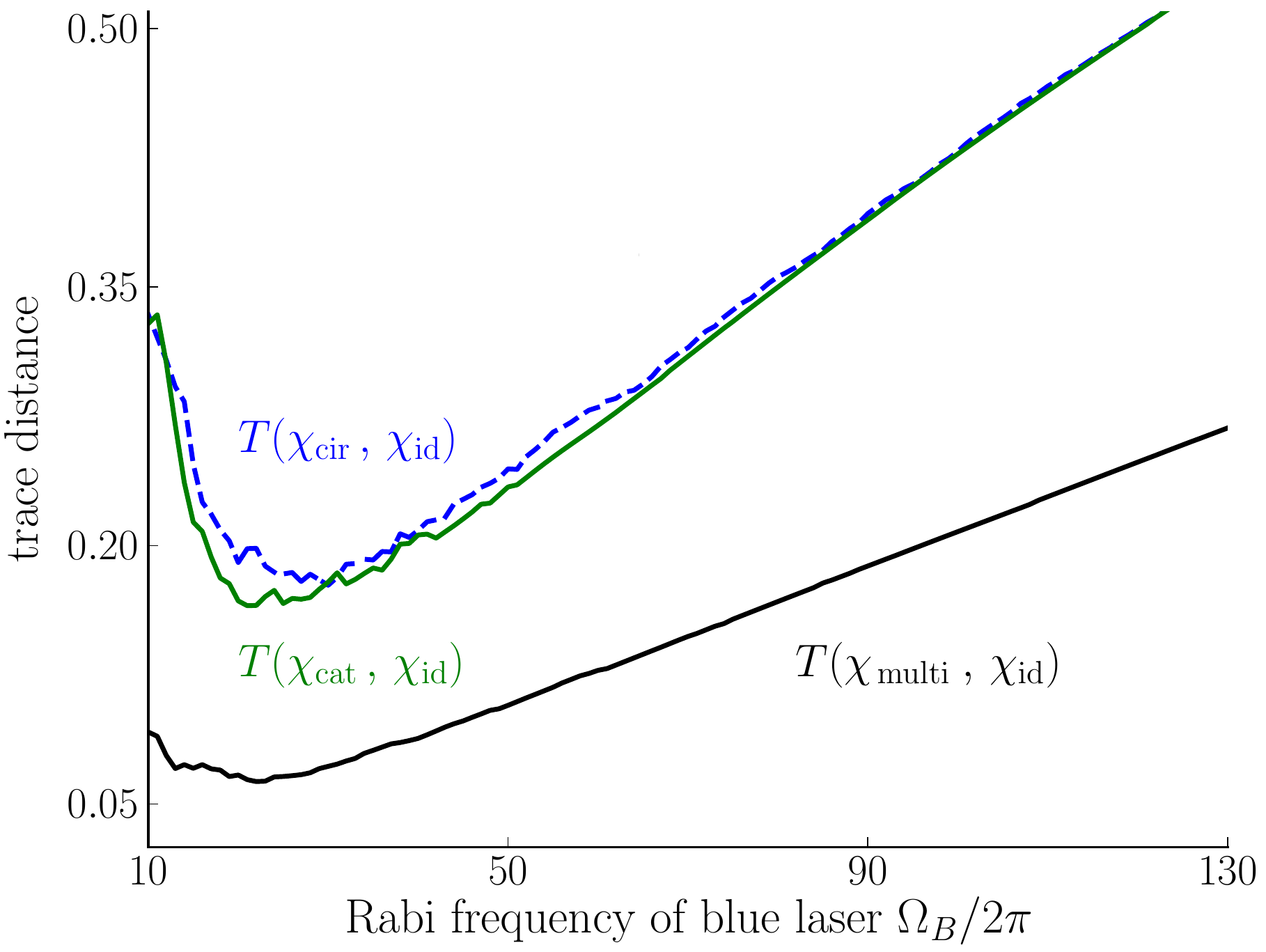}
\caption{\label{fig:toff} Trace distance between the process matrix $\chi_{\rm{id}}$ for the ideal Toffoli gate and the process matrices for Rydberg interaction implementations subject to dissipation and decoherence. The results are shown as a function of the Rabi frequency $\Omega_B$ of the $|p\rangle \rightarrow |r\rangle$ (blue) laser coupling. The dashed curve is obtained by simulating all three qubits as they evolve under the sequence of gates in the Toffoli circuit shown in Fig.~\ref{fig:toffcircuit}(a). The top solid curve uses concatenation of the one- and two-qubit process matrices to compute the Toffoli circuit process matrix. The bottom (solid) curve results from simulating the multi-qubit implementation shown in Fig.~\ref{fig:cknot}. In all calculations 500 Monte Carlo trajectories were used with the parameters listed in Table~\ref{tab:params}. }
\end{figure}

We demonstrate the characterization of the Toffoli gate resulting from simulation in Fig.~\ref{fig:toff}. The process matrix $\chi_{\rm{cat}}$ of the Toffoli gate in the circuit implementation (Fig.~\ref{fig:toffcircuit}) may be obtained without further simulation by a concatenation of the single qubit process matrices and the C-NOT process matrix of Sec.~\ref{sec:rydgte}A. Alternatively, we may simulate the circuit implementation in the full three-qubit Hilbert space to obtain $\chi_{\rm{cir}}$. In the simulation of a Rydberg mediated gate, the characterization of a single qubit has a Hilbert space dimension of $d=4^2$, which translates into a $4^6$ problem for the three qubit circuit characterization.

The top dashed (solid) curve in Fig.~\ref{fig:toff} illustrates trace distance between the full circuit $\chi_{\rm{cir}}$ (concatenated $\chi_{\rm{cat}}$) process matrix to the ideal process matrix $\chi_{\rm{id}}$, plotted as a function of $\Omega_B$. Each point in both curves is determined by propagating 500 wave function trajectories. The discrepancy between the two curves is due to $\chi_{\rm{cat}}$, which makes use of the same simulated C-NOT process matrix several times. Concatenating the process matrices of the same C-NOT simulation also ``concatenates'' the error associated with the simulation. The total error grows (nonlinearly) with the number of gates. This is demonstrated in Fig.~\ref{fig:toff}, where the discrepancy between the curves depends on $\Omega_B$. For lower values of $\Omega_B$, and thus slower gate operation, gate error is dominated by (non-unitary) processes such as dephasing. As correct Monte Carlo wave function statistics are sensitive to non-unitary errors, we find at small $\Omega_B$ the relatively large discrepancy between ${\rm Tr}(\chi_{\rm cat},\chi_{\rm id})$ and ${\rm Tr}(\chi_{\rm cir},\chi_{\rm id})$ to be expected at slow gate speeds (see Ref.~\cite{PhysRevA.54.5275} for an analysis of a similar situation). This is in contrast to results at high values of $\Omega_B$, where gate error is unitary and caused by an imperfect blockade. The discrepancy between ${\rm Tr}(\chi_{\rm cat},\chi_{\rm id})$ and ${\rm Tr}(\chi_{\rm cir},\chi_{\rm id})$ vanishes as $\Omega_B$ increases. We note that the Toffoli gate consists of six C-NOT gates and the trace distance to the ideal gate is, indeed, roughly six times the one shown in Fig.~\ref{fig:cnot}(c).

\begin{figure}
\includegraphics[width=0.82\columnwidth]{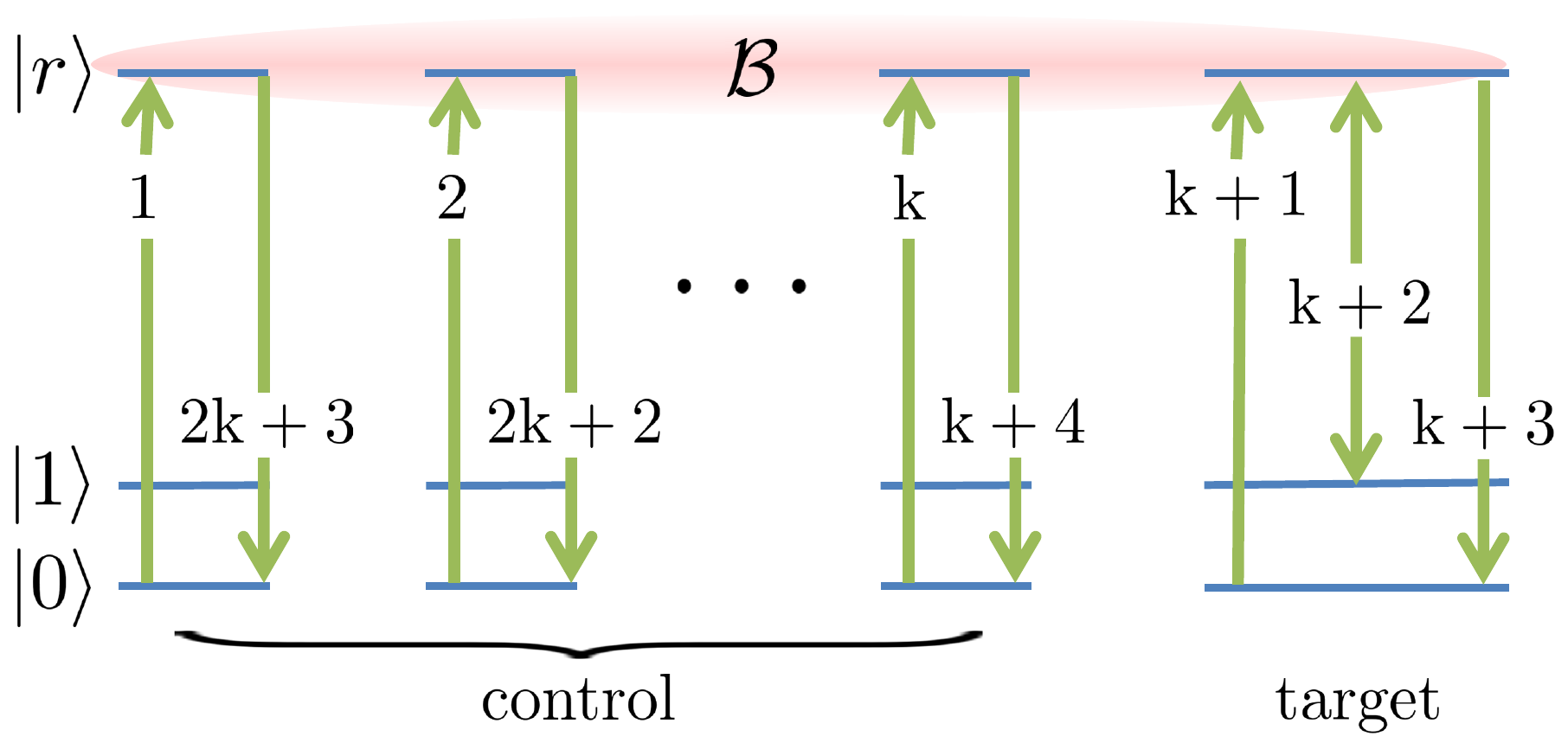}
\caption{\label{fig:cknot} Multi-qubit Rydberg blockade implementation of a C$_k$NOT gate. Each control atom is sequentially excited from $|0\rangle$  to $|r\rangle$ in $k$ $\pi$-pulses. Next, the target atom qubit states are swapped as in Fig~\ref{fig:cnot}(b), via the Rydberg state $|r\rangle$ (conditioned on no control atom populating the state $|r\rangle$). The control atoms are then returned to their original state in reverse order. The trace distance between the process matrix using this implementation for $k=2$ and the ideal Toffoli gate is shown as the lower curve in Fig.~\ref{fig:toff}.}
\end{figure}

A Rydberg excited atom blocks excitation of any number of atoms within the Rydberg interaction blockade radius, which may be of order 10 $\mu$m. Thus, it is possible to contain an entire qubit register within a single blockade radius, allowing implementation of multi-qubit gate operations which are faster than the circuit equivalent~\cite{molmer2011efficient}. One such protocol is the C$_k$-NOT gate operation, illustrated in Fig.~\ref{fig:cknot}~\cite{cknot}.

For $k=2$ the gate becomes the Toffoli gate and calculation of the process matrix is only possible by solving the master equation for the complete qubit register. In this paper, simulation of the process, including the decay and decoherence mechanisms detailed above, was carried out using the Monte Carlo method. The trace distance between the process matrix resulting from simulation and the ideal process matrix is shown as the lower, black curve in Fig.~\ref{fig:toff}. Remarkably, the multi-qubit implementation, with interactions allowed between all three atoms, performs markedly better than the Toffoli circuit consisting of one- and two-qubit operations. In comparison with the C-NOT gate, the minimal trace distance here is approximately 1.5 times larger. This is consistent with using 7 $\pi$-pulses rather than the 5 needed for a single C-NOT gate.

\section{\label{sec:DC}Conclusion}
In conclusion, we have presented an efficient method to compute the accumulation of errors in quantum circuits comprised of several few-qubit gates. Assuming the independence of errors over time and qubit register location we have shown that a set of concatenation rules on the appropriate few-qubit gate process matrices is enough to reproduce the process matrix of the entire circuit. To demonstrate the method's efficiency at calculating process matrices of large systems we considered the three-qubit Toffoli gate. The Toffoli gate may be implemented as a circuit of one- and two-qubit gates and simulations show that the process matrix obtained via concatenation is in good agreement with the result achieved by propagation through the entire circuit.

Our theory allows comparison of different implementations of gates. In particular, we compared a multi-qubit implementation of the Toffli gate with its one- and two-qubit circuit implementation. For the parameters chosen, the factor determining gate fidelity was the number of laser $\pi$-pulses. More gates lead to a lower fidelity, with a dependence that is almost linear. In this way, our analysis provides the necessary information to choose between different gate implementations. A theory of full error correction may benefit significantly from knowledge of the precise nature of errors incurred, potentially leading to higher thresholds for errors that can be remedied by appropriate error correction. The full process matrix, which remains at our disposal, may be further applied to optimally combine the Toffli gate with previous and subsequent gate operations along the lines of NMR composite pulses~\cite{Levitt198661}.


\begin{acknowledgments}
The authors gratefully acknowledge discussion with Mark Saffman. This work was supported by the IARPA MQCO program through ARO contract W911NF-10-1-0347 and by the Villum Foundation.
\end{acknowledgments}

\input{concat5.bbl}

\end{document}

%% file: concat5.bbl
%